\documentclass[12pt]{article}
\usepackage{indentfirst}
\usepackage{mathpazo}
\usepackage{setspace}
\usepackage{rotating}
\usepackage{bbm}
\usepackage{graphicx}
\usepackage{tikz}
\usetikzlibrary{decorations.pathreplacing,calligraphy}
\usepackage{rotating}
\usepackage{amsmath,amsthm}
\usepackage{amssymb}
\usepackage{booktabs} 
\usepackage{fullpage}
\usepackage{multicol}
\usepackage{lettrine}
\usepackage{bm}
\usepackage{algorithm}
\usepackage{algorithmic}

\usepackage[bottom]{footmisc}
\usepackage{tablefootnote}

\usepackage{float}
\usepackage{threeparttable}

\newtheorem{theorem}{Theorem}

\theoremstyle{definition}

\theoremstyle{example}

\newcommand{\E}{\mathbb{E}}
\usepackage[colorlinks=true,
citecolor=blue,
linkcolor=blue,
anchorcolor=blue,
urlcolor=blue]{hyperref}

\usepackage[authoryear]{natbib}
\usepackage[titletoc,toc,title,page,header]{appendix}
\usepackage{minitoc}

\noptcrule
\usepackage{caption}
\usepackage{subcaption}

\title{Quantifying How Much Has Been Learned from a Research Study\thanks{Acknowledgements: We thank Dave Blei, Macartan Humphreys, and Thomas Leavitt for conversations that inspired this project.  We are grateful to Kevin Esterling, Jiawei Fu, and Andrew Little for their helpful feedback on early drafts. This research was funded by internal support from Columbia University.  All data and code used to produce this paper is available on Github at \href{https://github.com/JonasMikhaeil/QuantifyingLearning}{https://github.com/JonasMikhaeil/QuantifyingLearning}.}}

\author{Jonas M. Mikhaeil\footnote{Ph.D. Student in the Department of Statistics, Columbia University \url{jmm2506@columbia.edu}}
\quad Donald P. Green\footnote{Burgess Professor of Political Science, Columbia University \url{dpg2110@columbia.edu}}
}

\begin{document}
\maketitle

\thispagestyle{empty}
\singlespacing

\begin{abstract}
How much does a research study contribute to a scientific literature? We propose a learning metric to quantify how much a research community learns from a given study. To do so, we adopt a Bayesian perspective and assess changes in the community’s beliefs once updated with a new study's evidence. We recommend the Wasserstein-2 distance as a way to describe how the research community's prior beliefs change to incorporate a study's findings. We illustrate this approach through stylized examples and empirical applications, showing how it differs from more traditional evaluative standards, such as statistical significance. We then extend the framework to the prospective setting, offering a way for decision-makers to evaluate the expected amount of learning from a proposed study. While assessments about what has or could be learned from a research program are often expressed informally, our learning metric provides a principled tool for judging scientific contributions. By formalizing these judgments, our measure has the potential to allow for more transparent assessments of past and prospective research contributions.
\end{abstract}

\section{Introduction}
Many important decision-makers in the research community -- funders, journal editors, promotion committees -- routinely make assessments of what has been learned, and what could be learned, from a research program.
In the social sciences, such assessments are often developed and expressed informally, characterizing in general terms how thinking on a given topic evolved in the wake of a specific study or, for research proposals, how they have the potential to shed light on outstanding empirical or theoretical questions.


In this paper, we attempt to formalize how much a study has contributed or could contribute to what is known about a quantity of interest, such as the effect of a policy intervention.  We take a Bayesian perspective and quantify the difference between the prior distribution (i.e., what the research community thought prior to obtaining results from a given study) and the posterior distribution (i.e., what the research community thinks now that it has seen the study's results). 
We propose the Wasserstein-2 distance, which has the advantage of being intuitive and easy to calculate.  We provide some worked examples that illustrate how this learning metric deviates from other evaluative standards, such as statistical significance. We conclude by discussing how this metric may be applied prospectively to research proposals in order to characterize how much we expect to learn from a future study.


\section{Measuring the Amount of Learning}

We want to quantify how much a scientific community learns when its prior about a quantity of interest is updated to form a posterior.
Suppose our beliefs, which may be derived from a scientific literature, provide us with a prior $\pi(\theta)$ about a parameter $\theta$. We characterize this process in abstract terms, but in practice methods of prior elicitation \citep{OHagan_2006,Mikkola2023} may be used to extract priors from both the literature or directly from domain experts.

Now suppose that new evidence, say in the form of a new study $y$, updates the scientific community's beliefs. Given this new information, the prior is updated to a posterior $\pi(\theta | y)$. 
How can we quantify how much has been learned? 

\begin{figure}
    \centering
    \includegraphics[width=0.5\linewidth]{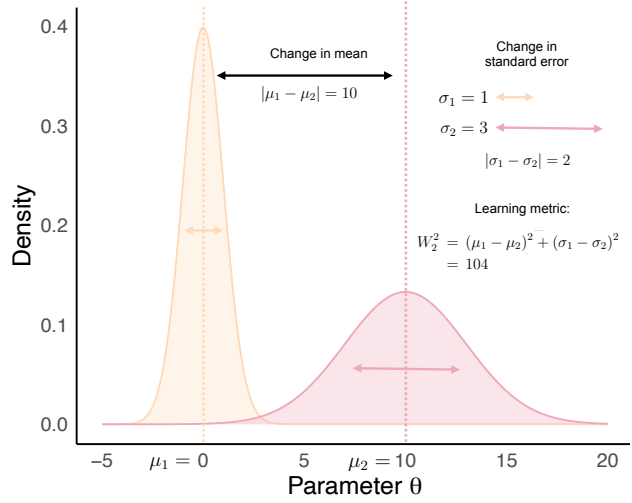}
    \caption{When beliefs change, both the magnitude (location) and the degree of uncertainty (scale) of a research community's posterior may change. Both kinds of changes factor into our learning metric.}
    \label{fig:MainIllustration}
\end{figure}

Figure \ref{fig:MainIllustration} illustrates this situation. A posterior encoding beliefs about a phenomenon under study may differ from its prior in crucial ways: Our belief about a parameter's magnitude may change. In this case there will be a shift in the posterior's location. 
Upon encountering new evidence, our uncertainty may also change -- if the new evidence confirms what was already believed, uncertainty decreases; if a surprising discovery is made, uncertainty may increase. These changes correspond to a change in the distribution's dispersion.

We recommend the Wasserstein distance \citep{Villani_2008} as an easy-to-compute measure of learning that is sensitive to both changes in our beliefs of an effect's magnitude and changes in our collective uncertainty. 
We introduce this measure in the next section (Section \ref{Sec:LearningMetric}) and provide a decision-theoretic justification for it in Appendix \ref{Sec:Theory}. In the Appendix (\ref{Sec:Lindley}, \ref{Sec:KL} and \ref{Sec:Comp}), we also compare our learning metric to other possible measures.

Notice that care has to be taken that the constructed prior represents our scientific beliefs. While in many Bayesian settings weakly informative priors yield similar posterior inference, when quantifying learning the exact specification of the prior matters. The amount learned for two priors that give similar posteriors may be drastically different.

\subsection{Wasserstein learning metric}
\label{Sec:LearningMetric}

 We now turn to the question of quantitatively characterizing how much a new study has changed our collective beliefs. 
To do so, we need to assess how much our prior $\pi(\theta)$ differs from the posterior $\pi(\theta | y)$ that has been updated based on the evidence $y$ of the new study.  
A natural choice to quantify the distance between the old prior $\pi_{old}:=\pi(\vartheta)$ and the new posterior $\pi_{new} := \pi(\vartheta | y)$ are the Wasserstein-$p$ distances
\begin{align}
\label{eq:LM} W_p(\pi_{old}, \pi_{new}) = \bigg(\inf_{\gamma \in \Pi(\pi_{old},\pi_{new})} \int \lvert x_1- x_2 \rvert^p d\gamma(x_1,x_2)\bigg)^{1/p}.
\end{align}
We give a decision-theoretic justification for the Wasserstein learning metric in Appendix \ref{Sec:Theory}.
The Wasserstein-p distances are easy to compute (especially when we are interested in posteriors of a one-dimensional parameter, such as the average causal effect of an intervention) and can be understood as the (p-th root) of the minimal cost of moving all the mass from the prior $\pi(\vartheta)$ to the posterior $\pi(\vartheta | y)$ \citep{Villani_2008}. 
When both the prior and posterior distributions are normal, the Wasserstein-$2$ distance has a closed-form solution and is readily interpretable:
\begin{align}
\label{eq:W2normal}
     W_2(\mathcal{N}(\mu_1,\sigma_1), \mathcal{N}(\mu_0,\sigma_0)) = \sqrt{(\mu_1 - \mu_0)^2 + (\sigma_1 - \sigma_0)^2},
\end{align}
\noindent where $\mu_0$ and $\sigma_0$ refer to the mean and standard deviation of the prior and where $\mu_1$ and $\sigma_1$ refer to the mean and standard deviation of the posterior. (We discuss how to calculate the Wasserstein-$p$ distance when normality is violated in Appendix \ref{App:Non-normal}.)
Notice that both changes in mean $\mu_j$ and changes in standard deviation $\sigma_j$ contribute to the Wasserstein distance. Learning is said to take place when the posterior shifts to the left or right on the number line because the metric does not presume that the truth is known.  Learning is also said to occur when new data cause the posterior to concentrate on a given location (a reduction in standard deviation) or when fresh information about the weaknesses of past research, such as revelations about fraud, introduce new uncertainty (an increase in standard deviation).


The Wasserstein learning metric is scale-dependent.  Care must be taken when interpreting a given value, as it reflects both the treatment dosage and the outcome metric. We recommend two comparisons to aid interpretation of a specific learning value. First, the Wasserstein learning metric can be divided by the prior standard error $\sigma_0$. This normalization is useful because a value of $1$ is what would be obtained when all uncertainty about the effect of interest is expunged while the mean of the distribution remains fixed. Second, researchers can and should compare learning values across studies within a given substantive domain.  When inputs and outcomes are scaled in the same way, within-domain comparisons give a sense of the studies' relative contributions \citep{learning_2025}.

\section{Applications}
\subsection{Illustrative Examples}
\label{Sec:StylizedExample}
\begin{figure}
    \centering
    \includegraphics[width=1\linewidth]{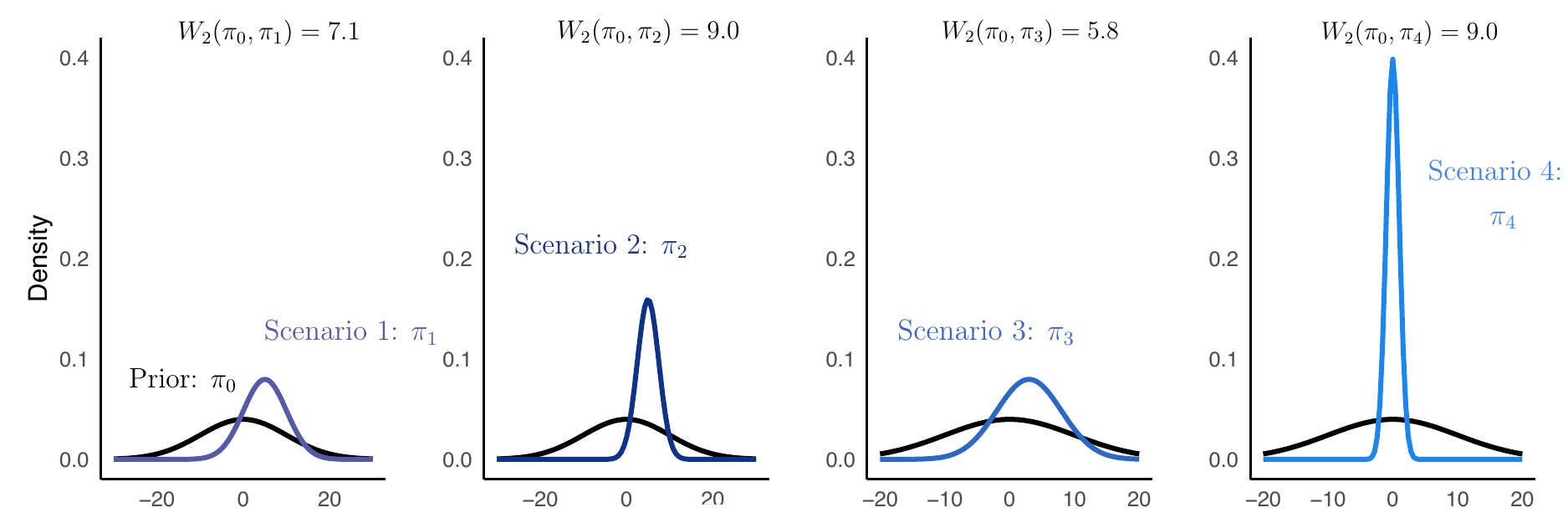}
    \caption{Comparison of four posterior distributions to a prior. How much is learned depends both on the change in location as well as the change in scale.}
    \label{fig:toy-example}
\end{figure}
In this section, we illustrate the use of the Wasserstein learning metric with a series of examples to guide intuition. Suppose we begin with a diffuse prior $\pi_0 = \mathcal{N}(0,10)$ centered around $0$ and with a standard deviation of $10$. We perform a study $y$, and our beliefs are updated accordingly.

Suppose that the study's result were an estimate of $6.67$ with a standard error of $5.77$.  And suppose that the study were large enough that the Central Limit Theorem could be invoked to argue that the sampling distribution were normal.  Using Bayes' Rule to update the prior based on the study's results yields a posterior that is $\pi_1=\mathcal{N}(5,5)$.   Applying the Wasserstein-$2$ distance to the comparison of a $\mathcal{N}(0,10)$ prior to a $\mathcal{N}(5,5)$ posterior yields a value of $7.1$.  In order to put that number in perspective, suppose instead that the study had yielded an estimate of $5.34$ with a standard error of $2.58$, so that the posterior were $\pi_2=\mathcal{N}(5,2.5)$. The posterior in Scenario 2 has a standard deviation that is half as large as the posterior in Scenario 1, and the Wasserstein-$2$ value grows to 9.0.  Comparing Scenario 1 to Scenario 2 illustrates the way the Wasserstein-2 distance recognizes an improvement in precision.

Suppose the study's results had instead come out such that the posterior distribution were $\pi_3=\mathcal{N}(3,5)$.  Call this Scenario 3. Compared to Scenario 1, Scenario 3 has the same posterior dispersion, but now the mean moves from $0$ to $3$ rather than from $0$ to $5$. The smaller shift in location is reflected in the Wasserstein-$2$ value of $5.8$.  

Finally, Scenario 4 considers the case in which the study generates estimates that render a posterior of $\pi_4=\mathcal{N}(0,1)$. In other words, the study leaves the prior location unchanged but greatly reduces the standard deviation.  
Comparing Scenarios 2 and 4 is telling: The same amount has been learned even though the location of the posterior shifted noticeably in Scenario 2 (it changed from $0$ to $5$) but not at all in Scenario 4. 

It should be stressed that the Wasserstein-$2$ characterization of what was learned from these four studies is only weakly related to conventional declarations about the statistical significance of each study's results.  
Although the posterior distributions in scenarios 1, 2, and 4 might be interpreted as insufficient to reject the null hypothesis that $\theta=0$ at the $0.05$ significance level, what we learn from these studies varies markedly.  Moreover, we learn as much from Scenario 4 as we do from Scenario 2, despite the fact that the former leads us to conclude that $\theta$ is close to zero.




\subsection{Empirical Example}
\begin{figure}
    \centering
    \includegraphics[width=1\linewidth]{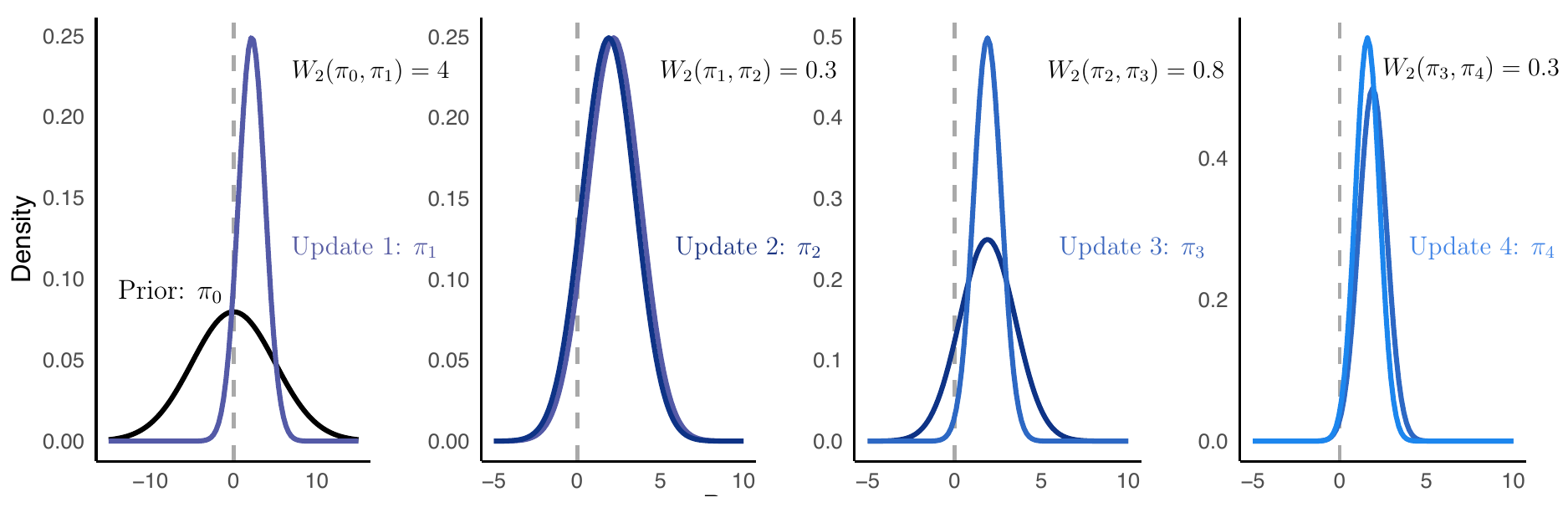}
    \caption{Comparison of the posteriors for the series of four field experiments presented in \citep{GREEN2016143}. In every panel, our learning metric $W_2$ gives a measure for how much the posterior has changed based on the new available evidence. }
    \label{fig:empirical_example}
\end{figure}

To illustrate how the Wasserstein learning metric may be applied to an actual research literature, we consider a series of four field experiments by \citet{GREEN2016143} that were designed to assess the extent to which lawn signs deployed shortly before elections raise the vote share of the advertising candidate. The four experiments, which were conducted in municipal, congressional, and statewide races, randomly assigned electoral districts (or precincts) to receive a few dozen roadside signs, while control districts received nothing. Outcomes were assessed by comparing candidates' official vote shares in treatment and control precincts. The estimates (standard errors) for the four studies, respectively, are 2.5 (1.7), -1.4 (5.7), 1.8 (0.9), and -1.2 (2.6).

Before this study, the effects of signage on candidates' vote share had seldom been the object of systematic evaluation.  Therefore, priors largely reflected the hunches of campaign experts. Judging from the tepid advice given by campaign professionals in their books about how to win campaigns, priors may be described as skeptical but diffuse.  Suppose, for example, that the prior distribution were $\pi_0 = \mathcal{N}(0, 5)$, implying that a roughly $95\%$ confidence set covers average treatment effects as low as -10 percentage points and as high as 10 percentage points. In Appendix \ref{Sec:Emp_ex}, we discuss another empirical example, a study investigating the economic effect of citizenship, where the prior is more ambiguous. 

The first study moves the prior $\mathcal{N}(0, 5)$ to a posterior $\mathcal{N}(2.2,1.6)$.  The resulting Wasserstein distance is therefore 
$\sqrt{2.2^2+(5-1.6)^2}=4.0$.  The second experiment moves the mean of the posterior but not its standard deviation: $\mathcal{N}(1.9, 1.6)$, which makes it particularly easy to calculate the Wasserstein distance of 0.3. 
The third experiment, conversely, moves the standard deviation of the posterior but not its mean: $\mathcal{N}(1.9, 0.8)$, which again makes it easy to calculate the Wasserstein distance of 0.8.  The final experiment results in a posterior of $ \mathcal{N}(1.6, .7)$, which implies a Wasserstein distance of 0.3.  In this case, the accumulation of all four experiments moves the prior $\mathcal{N}(0, 5)$ to a posterior of $ \mathcal{N}(1.6, .7)$, which implies a Wasserstein distance of 4.6. In this case, the distance between the starting point and the endpoint is somewhat smaller than the sum of the distances for each incremental study because the estimated experimental effects vary from study to study.





\subsection{Prospective Learning}
\begin{figure}
    \centering
    \includegraphics[width=1\linewidth]{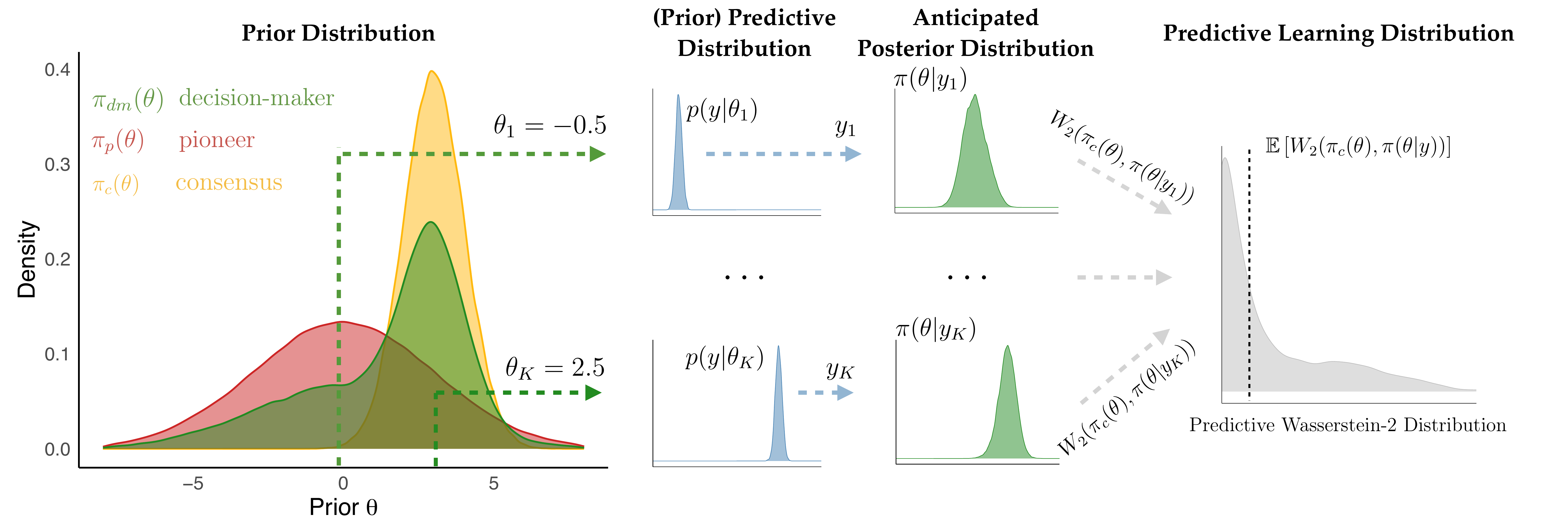}
    \caption{Illustration of prospective learning: Based on weighing the expert's judgment, a decision-maker holds a (decision-maker) prior $\pi_{dc}(\theta)$ as a mixture of the expert prior $\pi_p(\theta)$ and the consensus prior $\pi_c(\theta)$. This decision-maker prior implies a (prior) predictive distribution $p_{dc}(y)$ over data $y$. The decision-maker can then calculate the expected amount of learning by averaging the learning metric $W_2(\pi_c(\theta),\pi(\theta|y)$ over the predictive distribution $p_{dc}(y)$. }
    \label{fig:ProspectiveLearning}
\end{figure}

We will now venture from quantifying learning retrospectively to considering how study design and research portfolio allocation can benefit from considering how much may be learned from a future study. Both researchers and funders aim to maximize how much insight can be gained with the resources available. 
Decision-makers and grant-makers need to allocate limited resources across multiple possible research agendas. Researchers face design decisions such as tradeoffs between sample size and what population to draw subjects from. 
For example, it is typical for researchers to perform power calculations -- essentially asking how many experimental units are required to observe an anticipated effect with a prespecified power (often $80$\%) at a given level of confidence (often $95$\%). 
Instead of targeting significance, these decisions could be informed by maximizing how much we expect to learn from a study. 

How should we gauge prospective learning? The high-level idea is that our prior beliefs let us anticipate possible data. We can then calculate the expected learning for these anticipated datasets and average over how likely we think they are. This idea is illustrated in Figure \ref{fig:ProspectiveLearning}: Given a prior distribution on the parameters of the data generating process, we can calculate the (prior) predictive distribution $p(y) = \int p(y |\theta) d\theta$. Draws from this distribution serve as simulations for anticipated datasets.
Based on each of these envisioned datasets, we can update our prior to a posterior and calculate how much has been learned. 
Averaging over the (prior) predictive distribution $\mathbb{E}_{p(y)}\left[W_2(\pi(\theta), \pi(\theta|y) \right]$ then tells us how much we should expect to learn.

By way of a simple example, consider data drawn from a normal distribution $y \sim \mathcal{N}(\theta,\sigma)$ and let $\pi(\theta) = \mathcal{N}(\mu_{prior},\sigma_{prior})$ be our prior over the parameter of interest $\theta$. It is common for researchers planning a study to consult the literature to obtain a reasonable guess for the standard deviation of the observations $\sigma$. Assuming $\sigma$ and the effect size $\theta$ to be independent, we can update the prior to a posterior $p(\theta | y_1 , \dots y_n) = \text{normal}(\mu_{post},\sigma_{post})$, where
\begin{align}\nonumber
\mu_{post} = \frac{\frac{\mu_{prior}}{\sigma_{prior}^2} + \frac{n}{\sigma^2} \bar y}{\frac{1}{\sigma_{prior}^2}+\frac{n}{\sigma^2}}   \hspace{1cm} \text{and} \hspace{1cm} \sigma_{post}^2 = \frac{1}{\frac{1}{\sigma_{prior}^2}+\frac{n}{\sigma^2}}.
\end{align}
Notice that this posterior is random because of its dependence on $\bar y$. We can bound how much we should expect to learn from updating the prior $\pi(\theta)$ by taking the expectation of $W^2_2$ over the (prior) predictive distribution of $y_1,\dots,y_n$:
\begin{align} \nonumber
\mathbb{E}_{p(y_1,\dots,y_n)}\left[W_2(\pi_c(\theta), \pi_p(\theta | y))\right]^2 \, &\leq \, \mathbb{E}_{p(y_1,\dots,y_n)}\left[W^2_2(\pi_c(\theta), \pi_p(\theta | y))\right] \\ 
\, &= \, \mathbb{E}_{p(y_1,\dots,y_n)}\left[\left(\mu_{prior} - \mu_{post} \right)^2+\left(\sigma_{prior} - \sigma_{post} \right)^2 \right]\\ \nonumber
    \, &= \, 
    \frac{\sigma_{prior}^2}{\frac{\sigma^2}{n\sigma_{prior}^2}+1} +
    \sigma_{prior}^2 \left(1- \frac{1}{\left(1+\frac{\sigma_{prior}^2n}{\sigma^2}\right)^{1/2}}\right)^2.
\end{align}
Given our prior beliefs, there is no ex-ante reason to suspect the prior mean $\mu_{prior}$ and the posterior mean $\mu_{post}$ to vary systematically. 
The first term in the second line is due to random (and unsystematic) fluctuations expected in the posterior mean; with increasing sample size, this term converges to $\sigma_{prior}^2$ -- we do not expect to be too surprised by the anticipated data. The second term measures the increase in precision of the research community's beliefs about the parameter $\theta$. In the limit of large samples, this term converges to $\sigma_{prior}^2$ (all uncertainty is eliminated).

In order for our framework to  anticipate 
surprising empirical findings that challenge what we have previously believed, we need to introduce a crucial distinction:  When measuring how much has been learned, what matters is the change in beliefs of the relevant \textit{scientific community} at large -- not the change of an individual researcher's beliefs. Given extensive domain knowledge, a researcher might well anticipate an experimental outcome that for her peers comes as a surprise. While she may not have learned much, the scientific community did. We will refer to the community's prior as the \textit{consensus prior} $\pi_c(\theta)$, which comprises the beliefs that the relevant peers can agree on. It is this consensus prior that we can elicit from the literature, guided by a thorough and well-executed meta-analysis. We contrast this consensus prior with an expert's idiosyncratic prior. Based on theoretical or methodological insights, an expert may have good reasons to believe the past literature to be biased. This \textit{pioneer prior} $\pi_p(\theta)$ may motivate the researcher to plan a farsighted study to empirically support her suspicions.

Under this pioneer prior $\pi_p(\theta)$, the prior predictive distribution $p_p(y_1,\dots,y_n)$ can differ from the data that the larger community anticipates. 
Based on this envisioned data, the researcher can anticipate possible posteriors.
If she believes that her methodological critique will take hold and convince substantial portions of her research community, she may update her pioneer prior to model the anticipated posterior of the community. Otherwise, she needs to update the consensus prior with her anticipated data. With these posteriors $\pi(\theta|y_1,\dots,y_n)$ and her (prior) pioneer predictive distribution, she can calculate how much she expects the research community  to learn:
\begin{align}\nonumber
\mathbb{E}_{p_p(y_1,\dots,y_n)}\left[W_2(\pi_c(\theta), \pi(\theta|y_1,\dots,y_n) \right].
\end{align}

Under the pioneer prior, the posterior mean may be anticipated to systematically differ from the consensus prior mean. Additionally, posterior uncertainty may even be expected to increase beyond the level of the prior; the new results may cast doubt on what was previously believed to be known accurately. In these cases, researchers may expect to learn much even if their study is a pilot study with modest precision.

Depending on the expert's credibility and the soundness of her reasoning, decision-makers may accord her expectations more or less weight. They can do so by creating a decision-maker prior as a mixture of the consensus and pioneer prior. That is, $\pi_{dc}(\theta) = w \pi_p(\theta) + (1-w) \pi_{c}(\theta)$, where $w$ is the weight accorded to the pioneer prior. The decision-maker can then, similarly to the reasoning of the pioneer, anticipate a posterior and average how much is learned over their decision-maker predictive distribution $p_{dc}(y) = \int p(y|\theta) \pi_{dc}(\theta) d\theta $. Notice that the decision-maker, in their judgment of the pioneer's reasoning, already gauges how convincing they think the pioneer's reasoning is to the larger community. The decision-maker can thus calculate the anticipated posterior by updating their decision-maker prior. Decisions, such as the design of an experiment or which experiment to fund, can then be based on the expected degree of learning $\mathbb{E}_{p_{dc}(y_1,\dots,y_n)}\left[W_2(\pi_c(\theta),\pi(\theta|y_1,\dots,y_n)\right]$. Figure \ref{fig:ProspectiveLearning} illustrates this reasoning by which a decision-maker can quantify prospective learning. 

Let us consider the following simple example: Suppose we elicit a consensus prior  $\pi_c(\theta) = \mathcal{N}(3,1)$ from the literature -- the relevant scientific community holds that with relatively high certainty an effect is believed to be positive.
Based on careful methodological and substantive reasoning, an expert arrives at the conclusion that previous studies may have been biased and that there may be no effect after all. They hold the pioneer prior $\pi_p(\theta) =\mathcal{N}(0,3)$. This is the situation illustrated in the left panel of Figure \ref{fig:ProspectiveLearning}. 

\begin{figure}
    \centering
    \includegraphics[width=0.5\linewidth]{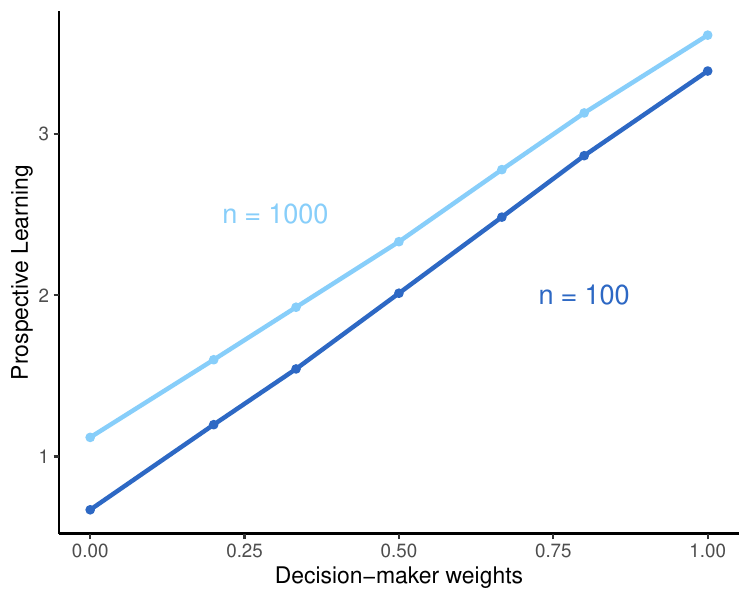}
    \caption{Degree of prospective learning for different study sizes $n$ as a function of the weight $w$ accorded to the pioneer prior $\pi_p(\theta)$.}
    \label{fig:weights}
\end{figure}
A decision-maker interested in funding the pioneer's study can gauge how much may be expected to be learned by weighing the pioneer and the consensus prior, that is, $\pi_{dc}(\theta) =w \mathcal{N}(0,3)+(1-w) \mathcal{N}(3,1)$.
Figure \ref{fig:weights} shows the degree of prospective learning the decision-maker can anticipate $\mathbb{E}_{p_{dc}(y)}\left[W_2(\pi_c(\theta),\pi(\theta|y)\right]$ for different sample-sizes of the proposed study as a function of the weight $w$ accorded to the pioneer prior.
When much weight is accorded to the expert's pioneering beliefs, much can be anticipated to be learned even from a pilot study of modest sample size. Note, however, that even if all weight is accorded to the consensus prior, we expect learning -- in this case we do not expect to be surprised, but we still expect future observations to decrease our uncertainty.

\section{Conclusion}

Off-the-shelf statistical tools have long existed to quantify the amount of learning that occurs in the wake of new evidence. Yet such metrics are seldom used for this purpose in practice; instead, assessments of research contributions are often informal and impressionistic.  This paper attempts to formalize the process of Bayesian updating so that one can characterize learning using a small number of inputs.  Indeed, the learning metric that we recommend, the Wasserstein-2 distance, is so simple in the case where the prior and posterior distributions are assumed to be normal that it can be calculated at a glance based on the means and standard deviations of the prior and the updated posterior.  For those cases in which it is unrealistic to assume normality of the prior or updated posterior, the Wasserstein-2 distance can still be calculated, albeit with a slightly more complex formula, as illustrated in Appendix \ref{App:Non-normal}.

This machinery allows researchers, journal editors, and funders to communicate in a more coherent and transparent fashion.  One might envision a summary, akin to the ``significance statement'' that many journals require, that depicts the prior, posterior, and Wasserstein learning score for a manuscript submitted for peer review. The prior could come from a formal meta-analysis, from a prediction market involving experts, or from an incentive-compatible elicitation survey \citep{Danz_2024}.
Granted, our learning metric is blind to the substantive importance of the research question itself and its units are specific to the scale of the outcome measure, but it nevertheless succinctly conveys the extent to which the study contributes to the research literature on a given topic.

Somewhat more complex is the problem of using this approach to characterize the value-added of a future study to a given research literature.  Here, the formal result hinges on whether a prospective study produces unexpected results, which in turn presupposes some difference of opinion within the research community --- if one set of scholars' expectations are vindicated by the new evidence, the other set of scholars will update markedly in the wake of findings that they find surprising. On the other hand, if everyone ex ante has the same priors, and the data come in as expected, the learning metric will simply reflect a decrease in uncertainty. Typically, the expected rate of learning will be greater when divisions of opinion imply that the anticipated study will produce results that will be surprising to some.

We close by acknowledging some of the limitations of the Wasserstein learning metric.  Unlike Lindley's learning metric (Appendix \ref{Sec:Lindley}), the Wasserstein distance is not cumulative.
That is, the amount learned from study to study according to the Wasserstein distance does not necessarily sum up to the Wasserstein distance between the initial prior and the final posterior.  When results fluctuate from one study to the next, as they do in our empirical example, the Wasserstein metric records the distance traveled along the way to the final destination.  While cumulativeness may be an appealing property, Lindley's learning metric comes with the important shortcoming of being insensitive to changes in the mean. As discussed in Appendix \ref{sec:perspLearning}, Lindley's measure is a compelling choice from the perspective of an oracle, whose posterior moves ever closer to the truth with every study. This oracle's perspective requires knowledge of each study's bias. Instead, the Wasserstein distance reflects how much posterior beliefs have changed in the wake of study after study, leaving open the possibility of being proven wrong about studies' biases.  In addition, the Wasserstein distance is specific to the units in which outcomes and treatments are scaled; this complicates the problem of comparing the amount of learning to be had from studies in different domains.  Nevertheless, for comparisons within a domain, this metric offers a convenient and concise summary of how much has been learned from new evidence.
\bibliographystyle{abbrvnat}
\bibliography{literature}

\appendix

\section{The Wasserstein Distance for Non-Normal Distributions}
\label{App:Non-normal}
\begin{figure}[th]
    \centering
    \includegraphics[width=0.7\linewidth]{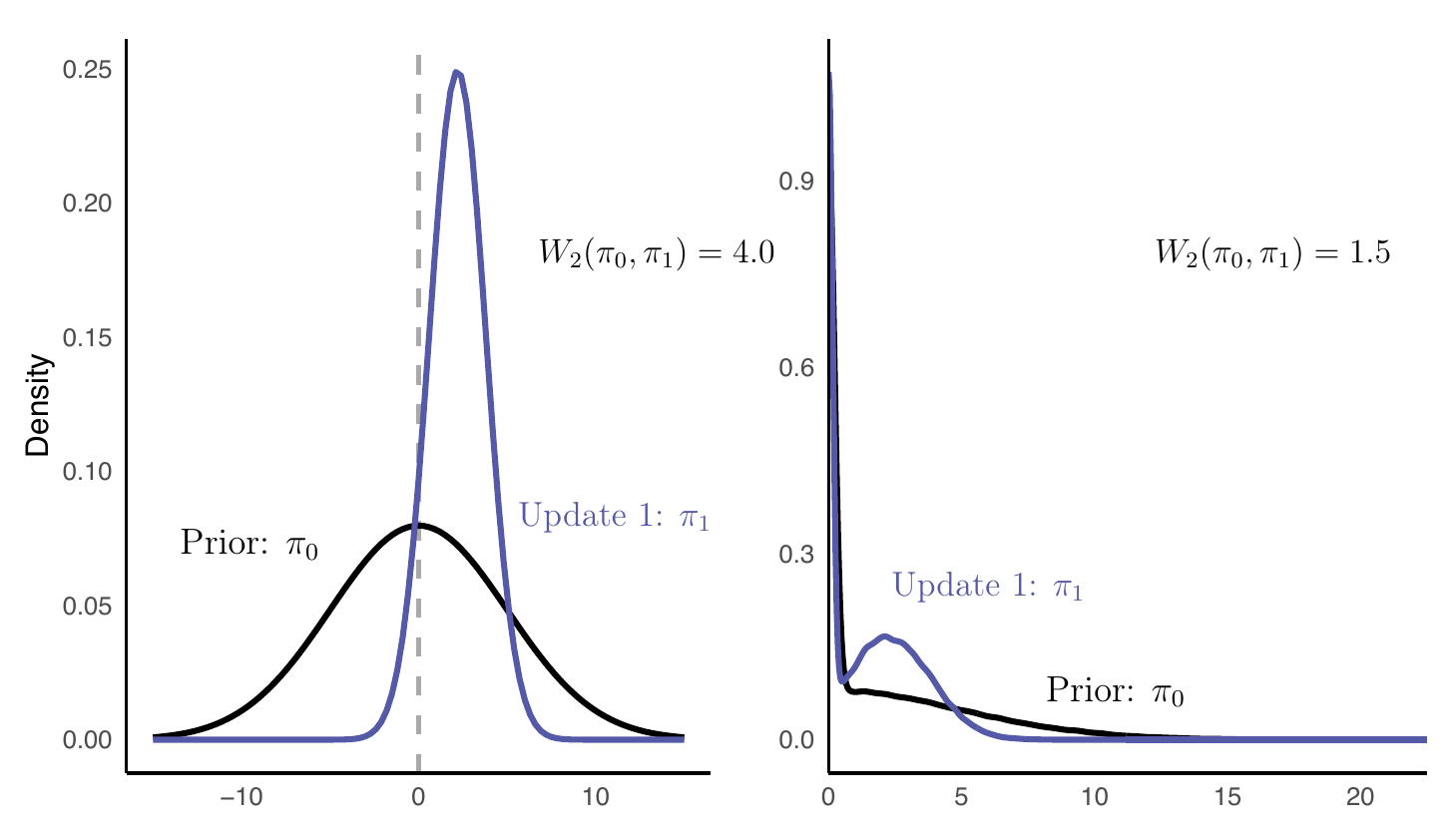}
    \caption{Comparison of posterior updating and learning in terms of our Wasserstein learning metric for the first field experiment presented in \citep{GREEN2016143} if the prior is chosen to incorporate the belief that the effect of interest is non-negative but probably small. The left facet shows how a normal prior is updated; the right facet, how a truncated normal is updated. }
    \label{fig:non-normal}
\end{figure}
In the main text, we proposed the Wasserstein distance as an attractive measure of how much has been learned when a prior is updated to a posterior. One of the benefits of this measure is its appealing closed-form solution in case both prior and posterior are normal, see Equation \ref{eq:W2normal}. The same closed-form solution obtains more general when prior and posterior are members of the same location-scale family \citep{Panaretos_19}. In these cases, practitioners can separately report the contribution of the change in mean and change in standard deviation to the overall amount of learning, see Table \ref{tab:decomp_table}.

In situations in which prior elicitation \cite{OHagan_2006,Mikkola2023} is used there may be justified worry that the elicited prior deviates significantly from normality. 
Fortunately, the Wasserstein-$p$ distances are straightforward to calculate for arbitrary (one-dimensional) distributions. Let $F$ be the cdf associated with the prior $\pi(\theta)$ and $G$ the cdf for the posterior $\pi(\theta|y)$, then 
\begin{align}\nonumber
    W^p_p(\pi(\theta) | \pi(\theta|y)) = \int_0^1 \lvert F^{-1}(t) - G^{-1}(t) \rvert^p dt,
\end{align}
where $F^{-1}(t)=\inf\{x \in \mathbb{R}:F(x)\geq t \}$ and $G^{-1}(t)=\inf\{x \in \mathbb{R}:G(x)\geq t \}$ are the generalized inverse functions of $F(t)$ and $G(t)$. 

In the case that prior and posterior are multivariate and non-normal, there is no general closed-form solution. The Wasserstein-$p$ distances can still be easily computed via linear programming. Off-the-shelf packages (such as the \href{https://cran.r-project.org/web/packages/transport/index.html}{transport} package in R \citep{transport}) make calculations of the Wasserstein distance straightforward. 

Figure \ref{fig:non-normal} shows an example in which the Wasserstein distance is used to quantify learning when prior and posterior are not normal. We return to the first field experiment presented in \citep{GREEN2016143} but update a non-standard prior reflecting the belief that the effect of lawn signs on voter turnout is non-negative but probably small. In this case, less is learned from the experiment because negative effects have been ruled out a priori.

For these settings with more complicated prior and posterior distributions, the amount learned cannot be neatly decomposed into contributions based on the change in mean and standard deviation. This makes sense given the fact that arbitrary distributions can change in ways not captured by their first two moments.
\begin{table}
    \centering
    \begin{tabular}{l|cccc}
     $N(0,10)$ &  $(\mu_1 - \mu_0)^2$ & $(\sigma_1 - \sigma_0)^2$  & $W_2^2$ &  $W_2$\\
        \hline
     $\rightarrow N(5,5)$  & 25  & 25 & 50 &  7.1\\
       $ \rightarrow N(5,2.5)$  &25 & 56.3  & 81.3& 9.0 \\
       $ \rightarrow N(3,5)$  & 49 & 25 & 74  & 5.8 \\
         $\rightarrow N(0,1)$ & 0 & 81 & 81 &9.0 \\

    \end{tabular}
    \caption{Contribution of the change in mean and the change in standard deviation to the Wasserstein-2 learning metric for the illustrative examples from section \ref{Sec:StylizedExample}.}
    \label{tab:decomp_table}
\end{table}
\section{A Decision-Theoretic Foundation for Learning Metrics}
\label{Sec:Theory}
To formalize how much a research community has learned from a study $y$, we aim to quantify how much their prior belief $\pi(\theta)$ has changed when updated to a posterior $\pi(\theta|y)$. Doing so involves making choices about which aspects of the distributions matter. We take a decision-theoretic perspective: The amount learned corresponds to how much our ability to make judgments about a decision has changed. While decision theory is largely concerned with making an optimal decision for a concrete situation, we argue that the yardstick by which to measure our beliefs is their ability to judge a decision for a wide variety of situations. In doing so, we essentially take a pragmatist perspective, judging our beliefs by their ability to solve problems \citep{James_07}. \footnote{To understand truth, James argues, we need to evaluate our beliefs considering the practical difference -- or the pragmatic ``cash-value'' \citep[1975, pg. 97]{James_07} -- that they bring about. What matters is how useful and dependable our beliefs are -- ``it is true because it is useful'' \citep[1975, pg. 93]{James_07}.
} This pragmatic view nicely fits with the well-known Bayesian standpoint that relates our degrees of belief with probabilities
and links those with how we ought to make decisions under uncertainty \citep{Ramsey_26,Finetti_37}.\footnote{The link between decision-making and probabilities is usually made in the context of gambling where deviations from probability calculus can be shown to lead to guaranteeable losses. Arguments of this kind, those that show agent's degrees of belief should satisfy the probability axioms, are referred to as Dutch Book \citep{Ramsey_26}.}

Learning, then, can be captured by the change in this ability to judge decisions.

Concretely, we ask how differently we would judge a decision $a$ if we base our expectation on the prior $\pi(\theta)$ instead of the posterior  $\pi(\theta|y)$. In decision theory, we judge a decision with respect to a utility (or loss) function $L(a,\theta)$ that describes the benefits and costs of a decision $a$ when the true state of the world is $\theta$. An essential part of knowledge is its generalizability. Hence, when quantifying how much has been learned, we aim not at making an optimal decision under a specific utility function $L(a,\theta)$, which would only capture the application of our knowledge in a specific circumstance but, instead, we ask how our judgments for a wide variety of reasonable situations and corresponding utility functions may change. 

Before seeing the new study, the community's beliefs are $\pi(\theta)$ and their expected utility from decision $a$ is given by $\E_{\pi(\theta)}[L(a,\theta)]$. This expectation changes to $\E_{\pi(\theta|y)}[L(a,\theta)]$ once the community updates its beliefs based on the new study $y$. The new study has thus changed our ability to make judgments by
\begin{align}\nonumber
    \left\lvert \E_{\pi(\theta)}[L(a,\theta)] - \E_{\pi(\theta|y)}[L(a,\theta)] \right\rvert.
\end{align}
The following theorem shows that changes in expected utility for a wide class of utility functions are continuous in the space of probability distributions endowed with the Wasserstein-$p$ metric. Concretely, we can guarantee small changes in how we judge an action when the prior $\pi(\theta)$ and the posterior $\pi(\theta|y)$ are close enough when measured by the Wasserstein-$p$ distance.
\begin{theorem}[Thm. 6.9 \citep{Villani_2008}]
\label{thm:WpCont}
    Let $\mu,\nu$ be measures in $\mathcal{P}_p(\mathbb{R})$ (i.e., with finite $p-$th moment), then for any $\varepsilon >0$, there exists a $\delta>0$ such that 
    \begin{align}
        W_p(\mu,\nu) \leq \delta \Rightarrow  \left \lvert \int \phi d\mu - \int \phi d\nu \right \rvert \leq \varepsilon
    \end{align}
    for all continuous function $\phi $ with restricted polynomial growth, that is, with $\lvert \phi(x) \rvert \leq C(1+|x-x_0|^p)$ for any $C,x_0\in \mathbb{R}$.
\end{theorem}
Let us illustrate the use of this theorem by way of a simple example. A common choice for a loss function is the quadratic loss $L_2(a,\theta) = \lvert a - \theta \rvert^2$. This loss is of restricted polynomial growth of order $2$. Theorem \ref{thm:WpCont} allows us to guarantee any level $\varepsilon > 0$ of consistency in judgment when our expectations pass from prior $\pi(\theta)$ to posterior $\pi(\theta|y)$ by requiring our beliefs before and after seeing the study to be sufficiently close in terms of the Wasserstein-$2$ distance. That is
\begin{align}
W_2\left(\pi(\theta),\pi(\theta|y)\right) \leq \delta \Rightarrow   \left\lvert \E_{\pi(\theta)}\left[  \lvert a - \theta \rvert^2 \right] - \E_{\pi(\theta|y)}\left[ \lvert a - \theta \rvert ^2\right]   \right\rvert \leq \varepsilon
\end{align}
for some $\delta>0$. We thus see that if the prior $\pi(\theta)$ is close to the posterior $\pi(\theta|y)$ in the Wasserstein-2 distance, we can guarantee that our evaluations of potential decisions $a$ will be very similar. 

Generally, learning will not be captured
by changes in expected utility for a specific utility function.
Our beliefs need to be useful and dependable across many situations.
The true appeal of the aforementioned theorem is that it shows that our Wasserstein learning metric allows us to guarantee small changes in our expectations for a wide range of possible utility functions. This is appealing because as a research community, we want to apply what we have learned to many different decision scenarios, each coming with its own utility function.  
A small value of our learning metric then implies that for a wide array of utility functions (and hence many situations) the new study would not lead us to judge decisions too differently, and we would conclude that little has been learned.  



\section{Perspectives on Learning}
\label{sec:perspLearning}
In this section, we briefly discuss possible perspectives on Bayesian learning. Lindley's seminal paper \citep{Lindley_1956} presents a natural starting point for our discussion (we present his measure in Appendix \ref{Sec:Lindley}). Lindley begins his discussion with seemingly reasonable requirements. Most importantly, the information provided by an experiment is additive. In our notation, if we were to move from a prior $\pi_0$ to an intermediate posterior $\pi_1$ to a second posterior $\pi_2$, then the amount of information $\mathcal{I}$ is supposed to add up. That is, the following ought to hold $\mathcal{I}(\pi_0,\pi_2) = \mathcal{I}(\pi_0,\pi_1) + \mathcal{I}(\pi_1,\pi_2) $. His measure is (up to multiplicative scaling and under a mild continuity assumption) unique in fulfilling this additivity property.

While this property may seem appealing, it comes at a cost. It tacitly assumes we are always homing in on the truth. Lindley's measure is the optimal choice if the model we use to update our posterior is correctly specified; in this case the posterior concentrates around the ``true'' effect size. 

Notice that in practice this assumption of a correctly specified model for posterior updating will rarely be plausible. When researchers update their beliefs, they do so based on the study at hand but also based on their experience and beliefs. These beliefs are fallible, however.
Not infrequently, scientific literatures experience upheavals when methodological critique show the previous literature to be biased.
Taking all of this into account, we cannot rely on the assumption of a correctly specified model for posterior updating; we need to be open to the possibility that assumptions that we formerly embraced will later be shown to be untenable.


Lindley's approach may be seen as one of two epistemological positions. Following Lindley, we can try to quantify learning in terms of the distance of our posterior from the ``truth.'' This is a cumulative view of learning; every study either moves us closer or further from knowledge. Doing so requires an oracle's perspective: We can only do so if we know the correct Bayesian model to update our beliefs -- this includes perfect knowledge about all studies' biases. If we are willing to consider a (currently) settled literature as the ``truth,'' we can measure retrospectively how far every study moved us toward or away from this current end-point of the literature.

Our paper presents a different, more pragmatic perspective. Acknowledging past upheavals of what was considered known, we need to be open to the possibility of being surprised yet again. Without an oracle's omniscience, what we hold as true at every given moment is described by our current posterior -- even if these beliefs may be invalidated in the future. Instead of measuring how much has been learned as the distance to an inaccessible ``true'' posterior, we suggest that learning is more fruitfully understood as changes in our beliefs per se. Whenever we update our posterior, something is learned. This pragmatic perspective on learning is inherently non-cumulative. Not positing a truth we move toward, nothing distinguishes moving from a prior $\pi_0$ to a posterior $\pi_1$ from moving into the opposite direction, that is, back from the posterior $\pi_1$ to the prior $\pi_0$. In both cases, an equal amount is learned. Moving back and forth, we travel twice the distance even if we end up where we started. 
This non-cumulative view contrasts with the cumulative viewpoint championed by Lindley from which we would have to judge a scientific community 
moving back to its original prior to have learned nothing in the process.
\section{Lindley's Learning Metric}
\label{Sec:Lindley}
In his seminal paper, Lindley \citep{Lindley_1956} develops a measure of how much information a study has provided: 
\begin{align}
    \mathcal{I}(\pi(\vartheta),y) \, &= \, H(\pi(\vartheta\lvert y)) - H(\pi(\vartheta)) \\ \nonumber
    \, &:= \,\mathbb{E}_{\pi(\vartheta\lvert y)}\bigg [ \log\pi(\vartheta\lvert y) \bigg] - \mathbb{E}_{\pi(\vartheta)}\bigg [ \log\pi(\vartheta) \bigg]
\end{align}
This expression depends on the observed data $y$; some experimental outcomes are more informative than others. In addition to its information-theoretic appeal, Lindley's measure $I(\pi(\vartheta),y)$ can also be understood as the reduction of risk corresponding to a decision problem where one has to decide which distribution to report about an unknown quantity $\vartheta$ \citep{Parmigiani_1994}. The utility function $U(\vartheta, \varphi) = \log \varphi(\vartheta)$ implied by Lindley's measure is appealing from this decision-theoretic perspective as it induces decision makers to report their current beliefs in the form of their posterior. It is also commonly used in the context of Bayesian experimental design \citep{Huan_24,rainforth_24}.

While theoretically appealing, Lindley's measure $I(\pi(\vartheta),y)$ can be hard to compute in practice \citep{Batu_Dasgupta_Kumar_Rubinfeld_2002}. 
Similar to the Wasserstein-2 distance, see Section \ref{Sec:LearningMetric}, Lindley's measure of information has a closed-form solution in the situation in which prior and posterior are normal. 
\begin{align}
    \mathcal{I}(\text{normal}(\mu_{post},\sigma_{post}),\text{normal}(\mu_{prior},\sigma_{prior})) = \log{\sigma_{post}}-\log{\sigma_{prior}}.
\end{align}
This reveals why Lindley's information measure is unappealing in practice. 
If we acknowledge the possibility for studies to be biased, we are not guaranteed that the posterior mean concentrates on the true value of $\vartheta$. New studies may reveal bias in previous studies and impact our shared beliefs drastically by changing the posterior mean.
 We thus recommend using the Wasserstein distances as learning metrics (Equation \ref{eq:LM}) as they are sensitive to comprehensive changes in the posterior distribution. See also the discussion in the previous section (Section \ref{sec:perspLearning}).
\section{Kullback-Leibler Divergence}
\label{Sec:KL}
The Kullback-Leibler divergence
\begin{align}\nonumber
      \text{KL}(\pi(\theta \lvert y) , \pi(\theta)) = \int \pi(\theta \lvert y) \log{\frac{\pi(\theta \lvert y)}{\pi(\theta)}} d\theta.
\end{align}
is frequently used to compare how similar two distributions are \citep{Nielsen_2020}. 
How much is learned in terms of this divergence is upper bounded by the surprisal of the study's outcome $S(y) = \frac{1}{p(y)}$, where $p(y) = \int p(y|\theta) \pi(\theta) d\theta$ is the prior predictive distribution.
The Kullback Leibler divergence is, in contrast to the Wasserstein-$p$ distances, not a metric on the space of distributions because it violates the triangle inequality. Similarly to the Wasserstein-$2$ distances, it affords a closed-form solution between two normal distributions:
\begin{equation}
    \text{KL}(\text{normal}(\mu_1,\sigma_1),\text{normal}(\mu_0,\sigma_0)) = \frac{1}{2}\left(\frac{(\mu_0-\mu_1)^2}{\sigma_0^2}+\frac{\sigma_1^2}{\sigma_0^2} - \ln{\frac{\sigma_1^2}{\sigma_0^2}-1}\right).
\end{equation}
We can see that in the case in which both the prior and posterior are normally distributed, the Kullback-Leibler divergence and the Wasserstein-$2$ distance share important features: both are sensitive to changes in the mean as well as changes in the distribution's standard deviation. A fundamental difference is that the Kullback-Leibler divergence measures change in terms of relative changes in the standard deviation (that is, changes in scale), whereas the Wasserstein-$2$ distance is based on absolute changes.

For general distributions, there is no direct relationship (one-to-one or monotonic) between the Kullback Leibler divergence and the Wasserstein-$2$ distance. Similarly to the Wasserstein-$p$ distance which metrizes weak convergence with additional convergence of the first $p-$ moments, the Kullback-Leibler divergence $\text{KL}(P,Q)$ metrizes weak convergence with the additional constraint that $P$ is absolutely continuous with respect to $Q$.  

While the Kullback-Leibler divergence may be appealing in situations in which the prior and posterior are normal and relative changes in scale are of more substantive interest than absolute changes, the Kullback-Leibler divergence lacks the general decision-theoretic foundation we have proposed for the Wasserstein-$p$ distances.

\section{Comparison of Candidate Learning Measures}
\label{Sec:Comp}
\begin{table}
    \centering
    \begin{tabular}{l|ccc}
     $N(0,10)$ &  $W_2$ & $\text{KL}_{sym}$  &Lindley $\mathcal{I}$ \\
        \hline
     $\rightarrow N(5,5)$  &  7.1& 1.75 &  0.69\\
       $ \rightarrow N(5,2.5)$  &9.0 & 9.16  & 1.37 \\
       $ \rightarrow N(3,5)$  &5.8 &1.35  & 0.69 \\
         $\rightarrow N(0,1)$ &9.0  & 49 & 2.3\\

    \end{tabular}
    \caption{Comparison of our learning metric $W_2$ with the symmetrized Kullback-Leibler divergence $\text{KL}_{sym}$ and Lindley's Information measure $\mathcal{I}$, using the illustrative examples from section \ref{Sec:StylizedExample}.}
    \label{tab:comp_table}
\end{table}
Table \ref{tab:comp_table} compares the three candidate learning measures we have discussed. The table shows that the Wasserstein-$2$ distance and the Kullback-Leibler divergence largely agree in this stylized example. This is to be expected as long as the prior and posterior are normal and there is no change in scale (i.e., as long as the relative and absolute changes in standard deviation remain comparable). Changes in scale can lead the Kullback-Leibler divergence and the Wasserstein-$2$ distance to disagree. While equally much is learned in scenario$_2$ and scenario$_4$ in terms of the Wasserstein-$2$ distance, the Kullback-Leibler divergence is sensitive to the change of scale (in scenario$_4$ the standard error changes by a factor of 10) and ascribes significantly more learning in this case.

Lindley's learning metric is unable to capture learning because it is completely insensitive to changes in the mean -- it, for example, ascribes the same amount of learning for scenario $1$ and scenario $3$ even though the mean has further from prior beliefs in the first example.

\section{Empirical Example: Measuring Learning when Priors are Ambiguous}
\label{Sec:Emp_ex}
To illustrate how one might gauge a study's contribution to a research literature that is experiencing a clash between competing research methods, we turn to a recent study conducted by \citet{Hainmueller_Cascardi_Hotard_Koslowski_Lawrence_Yasenov_Laitin_2023}, which uses a randomized control trial to investigate the economic effects of citizenship. One noteworthy feature of this study is that it is the first randomized field experiment to investigate whether acquiring citizenship increases subjects' income.  The researchers use an encouragement design that lowers the effective cost of obtaining citizenship for a population of legal residents in New York City whose financial records could be tracked for three years (N=1,454).\footnote{As the authors explain on p.4, ``Eligible lawful permanent residents could sign up for a lottery that randomly selected
some registrants to receive a voucher to pay for the naturalization application fee.''} Approximately 36 percent of subjects are ``compliers'' in the sense that they obtain citizenship if and only if encouraged by the experimental intervention.  Another noteworthy feature of this experiment is that, despite its size and statistical power, its estimate of the complier average causal effect (CACE) is close to zero.  In standardized terms, compliers' income increases by only 0.074 standard deviations in the wake of treatment, with a standard error of 0.121.\footnote{The estimated CACE on log income reported in panel B of Table 4 is 0.025 with a standard error of 0.045. The standard deviation at baseline is 0.34 according to Table 2.} From the vantage point of conventional hypothesis testing, the fact that the study did not find statistically significant income gains as a result of naturalization might be construed as a failure, insofar as the policy-minded reader reverts to the null hypothesis of no effect. Our learning metrics, however, suggest that much has been learned from this study.

In order to appreciate the study's contribution, consider the prior literature. Focusing on the most methodologically rigorous prior studies, one would come away with the impression that citizenship leads to increased income. Granted, there are nagging concerns about selection bias in the extant non-experimental literature (i.e., residents with higher incomes are especially likely to apply for citizenship) and evidence that observational estimates of the income effect dissipate when well-known correlates of income are controlled statistically \citep{DeVoretz_Pivnenko_2005,Engdahl_2014,Scott_2008,Bevelander_2006}.  On the other hand, some of the more credible quasi-experimental research designs show substantial income effects \citep{Helgertz_14,Corluy_11,Bevelander_12,Catron_19,Picot_11,Hainmueller_19}, and theoretical mechanisms for such a causal effect abound \citep{Hainmueller_19}. Based on the literature reviewed in the appendix of \citet{Hainmueller_Cascardi_Hotard_Koslowski_Lawrence_Yasenov_Laitin_2023}, a stylized description of the research literature preceding 
\citep{Hainmueller_Cascardi_Hotard_Koslowski_Lawrence_Yasenov_Laitin_2023}
is that it is a mixture of optimism and pessimism. The causal effect is widely suspected to be positive given the many reasons to suppose that in the long run citizenship boosts income; on the other hand, the threat of (positive) selection bias looms over much of the nonexperimental literature. 


For the sake of illustration, let's consider two readings; in both  \citet{Hainmueller_Cascardi_Hotard_Koslowski_Lawrence_Yasenov_Laitin_2023} study's finding of $\mathcal{N}(0.074,0.121)$ leads to considerable updating.  One reading expects citizenship to have a large positive effect but accords a great deal of uncertainty to the observational research designs and declares the prior to be diffuse: $\pi_{\text{old}1} = \mathcal{N}(0.3,0.3)$.  
\citet{Hainmueller_Cascardi_Hotard_Koslowski_Lawrence_Yasenov_Laitin_2023} study's finding updates this prior to a posterior $\pi_{\text{new}1}= \mathcal{N}(0.11, 0.11)$. Both prior and posterior are normal, such that we can use Equation \ref{eq:W2normal} to straightforwardly calculate our learning metric:
\begin{align}\nonumber
    W_2(\pi_{\text{old}1},\pi_{\text{new}1}) = \sqrt{\left(0.3 - 0.11\right)^2 + \left(0.3 - 0.11\right)^2} = 0.27.
\end{align}

A second reading takes a more skeptical position on average, on the grounds that some observational studies find results suggesting that citizenship has no effect. If we imagine that the prior distribution is a truncated normal --- with no mass below zero on the grounds that no one believes that citizenship diminishes earnings --- but with a right tail because there is reason to suspect that the effect might be large.  This second reading departs from the normal-normal formulation that underlies the W-2 distance metric, which gives us an opportunity to show how one can nevertheless apply the Wasserstein metric, using a more complex formula (we discuss how to calculate the Wasserstein distance for general distributions in Appendix \ref{App:Non-normal}).  
Imagine that 31\%  of the normal is truncated at zero and that rest spills into positive values with a mean of 0.2 and a standard deviation of 0.4.
Denote this prior with $\pi_{\text{old}2}$. Using computational methods, such as importance sampling, we can update this (non-normal) prior with the result from the \citet{Hainmueller_Cascardi_Hotard_Koslowski_Lawrence_Yasenov_Laitin_2023} study to obtain a posterior. Denote this posterior with $\pi_{\text{new}2}$ -- due to the non-normality of the prior, the posterior will, in general, also be non-normal.
In this case, we find
\begin{align}\nonumber
    W_2(\pi_{\text{old}2},\pi_{\text{new}2}) = 0.34
\end{align}

In both scenarios, we find that the ``null findings'' reported by \citet{Hainmueller_Cascardi_Hotard_Koslowski_Lawrence_Yasenov_Laitin_2023} lead to considerable updating.  Notice that because their study contributed to an existing literature, the priors were far from ``flat.''  One interesting feature of the Wasserstein metric is that the first study to speak to a research question tends to produce large gains in knowledge.  For example, if one were to begin with a prior of $\mathcal{N}(0,100)$, even a small study that yielded a posterior of $\mathcal{N}(0,19)$ might be said to have made a substantial contribution ($W_2=9$), even larger than the next study that moves the prior to $\mathcal{N}(0,3)$, such that ($W_2=4$).  When characterizing ``flat'' priors, however, it important to use some kind of systematic or incentive compatible method so that the priors are authentic.  Rarely is a literature so devoid of evidence or theory that the standard deviation is extremely large.
\end{document}